# Frequency up-conversion of an infrared image via a flat-top pump beam

CHEN YANG,[1, 2,*] SHI-LONG LIU,[1,2,*] ZHI-YUAN ZHOU,[1, 2, 3 †] YAN LI,[1, 2] YIN-HAI LI,[1,2,3] SHI-KAI LIU,[1, 2] ZHAO-HUAI XU,[1, 2] GUANG-CAN GUO, [1,2] AND BAO-SEN SHI [1, 2,3]

[1] CAS Key Laboratory of Quantum Information, USTC, Hefei, Anhui 230026, China
[2] Synergetic Innovation Center of Quantum Information & Quantum Physics, University of Science and Technology of China, Hefei, Anhui 230026, China
[3] Wang Da-Heng Collaborative Innovation Center for Science of Quantum Manipulation & Control, Heilongjiang Province & Harbin University of Science and Technology, Harbin 150080, China
* These two authors contributed equally to this work.
† Corresponding authors: zyzhouphy@ustc.edu.cn

**Abstract:** The infrared imaging detection is an important and promising technology having wide applications. In this work, we report on the frequency up-conversion detection of an image based on sum frequency generation in a nonlinear crystal with a flat-top beam acting as the pump instead of a Gaussian beam, the up-converted image at 525 nm falls in the sensitive band of visible detectors and human eyes. Both theoretical simulations and the experimental results clearly demonstrate that using a flat-top beam as a pump can improve the fidelity of an image after the up-conversion compared with a Gaussian pump beam. Our scheme will be very promising for infrared image detection based on frequency up-conversion.

Optical imaging systems using an infrared illumination have wide applications in many fields including biomedicine, remote sensing, and LIDAR applications [1–5]. The imaging system working at wavelength range around 1550 nm is of specific interest due to several advantages, including the facts that it works in an eye-safe spectral region and in a good atmospheric transparency window. Besides, there also exists available high peak power lasers in this range used as a pump for aching high conversion efficiency. However, CCDs working in this spectrum, commonly based on InGaAs materials, suffer from low efficiency and slow speed, high readout noise, and also more rigorous cooling requirements compared with the silicon-based CCDs working in the visible range [6]. Fortunately, imaging technology based on optical frequency up-conversion provides a promising way to visualize infrared images with standard silicon-based CCDs for taking advantage of the better performances, such as high efficiency, good resolution and easy operation without cooling, etc.

Frequency up-conversion based on a second-order nonlinear optical crystal, especially a period poled crystal for quasi-phase matching, is being investigated in recent years for the advantage of high efficiency [7-12]. Recently the sensitivity of imaging detection based on up-conversion obtained with a broadband pump is close to that obtained by the direct detection with CCD working in the infrared spectrum, and it can be further improved using a higher power laser [9]. This makes the infrared imaging up-conversion technology applicable in practice.

In addition to improving the conversion efficiency, the imaging quality is also an important issue in imaging up-conversion. It is based on sum-frequency generation, so the features of the pump beam crucially affect the quality of the imaging. In the previous works [7-12], a Gaussian beam is widely used as a pump, making the up-converted light have a Gaussian modulation factor. In this work, we use a focused beam with a flat-top intensity profile as a pump instead of a Gaussian beam to eliminate the intensity modulation. Such a flat-top beam can uniformly illuminate a specific volume of space that has a flattened irradiance profile with a uniform central region and can propagate through the working volume with small distortion of its uniformity [13]. The flat-top intensity profile is created in focal plane of a focusing lens when the input beam has Airy-disk intensity distribution, which can be easily created by a beam shaper [14].

In our up-conversion image scheme, we place the crystal in the image plane. The real image of the object is converted in the crystal and then the up-converted image in the visible range is imaged to the CCD chip by a lens. We first study the imaging system in theory and demonstrate the flat-top beam can equalize the intensity of the image as it should be and then perform two types of image up-conversion in the experiment to show the significant advantages in the case with a flat-top beam pump.

In the crystal, the infrared light is partially converted to the visible light, and the process is governed by the coupled wave equations shown in Eq. (1), which have been simplified using the paraxial approximation, slowly varying amplitude field approximation and plane wave approximation [15,16],

$$\frac{\partial E_1}{\partial z} = \frac{i}{2k_1}\nabla_\perp^2 E_1 + \frac{2i\omega_3 d_{eff}}{n_3 c} E_3 E_2^* e^{-i\Delta kz}$$
$$\frac{\partial E_2}{\partial z} = \frac{i}{2k_2}\nabla_\perp^2 E_2 + \frac{2i\omega_3 d_{eff}}{n_3 c} E_3 E_1^* e^{-i\Delta kz} \quad (1)$$
$$\frac{\partial E_3}{\partial z} = \frac{i}{2k_3}\nabla_\perp^2 E_3 + \frac{2i\omega_3 d_{eff}}{n_3 c} E_2 E_1 e^{i\Delta kz}$$

Here $E_i(x,y,z)$ represent three waves propagating in the crystal. The subscript $i = \{1, 2, 3\}$ represents infrared signal, pump, and up-converted light beam, respectively. The constant $\omega$, $n$, $c$ and $d_{eff}$ is angular frequency, refractive index, speed of light and effective nonlinear coefficient, respectively. $\Delta k$ represents the phase mismatching, which is generally set to be 0 for efficient conversion. In fact, the pump beam power is much stronger than that of the up-converted beam. We therefore only consider the third equation in Eq. (1).

In theory, the real image is only clear in the image plane. In reality, however, the real image is clear in a short range where the resolution is tolerable. Thus, in the up-conversion process, the crystal should be thin in order to obtain a clear up-converted image. In addition, the pump beam cannot be tightly focused because the width of the pump beam must be larger than the image in crystal, therefore the width of the pump field changes slowly in the propagation direction. Due to the short interaction distance, the solution of the third equation in Eq. (1) is approximately

$$E_3(x,y,z_c) \approx \frac{2i\omega_3 d_{eff} L}{n_3 c} E_2(x,y,z_c) E_1(x,y,z_c), \quad (2)$$

where $L$ is the length of the crystal, and $z_c$ is the coordinate of the crystal. We assume that the magnification rates of the two lens imaging system are $M_1$ and $M_2$ respectively, which are determined by the distances among the object planes, lenses and the image planes. According to the geometrical optics, the object-image relations are $E_1(x,y,z_c) = \frac{1}{M_1} E_1\left(\frac{-x}{M_1}, \frac{-y}{M_1}, z_o\right)$ and $E_3(x,y,z_i) = \frac{1}{M_2} E_3\left(\frac{-x}{M_2}, \frac{-y}{M_2}, z_c\right)$, where $z_o$ and $z_i$ are the coordinates of the object and CCD chip. Given the Eq. (2) and the intensity-field relation of $I \propto |E|^2$, we get the final object-image relation of the whole system

$$I_3(x, y, z_i) = KI_2\left(\frac{-x}{M_2}, \frac{-y}{M_2}, z_c\right) I_1\left(\frac{x}{M_1 M_2}, \frac{y}{M_1 M_2}, z_o\right) \quad (3)$$

Here $K = \left(\frac{2\omega_3 d_{eff} L}{n_3 c M_1 M_2}\right)^2$ is a constant factor. The Eq. (3) indicates that the image is modulated by the intensity distribution of the pump beam. To obtain an up-converted image with high fidelity, $I_2$ should be a constant in the crystal, which can be approximately realized using a flat-top beam mentioned before.

We then provide an accurate numerical simulation according to the method described in [12]. We choose a simple pattern formed by three rectangles as an object shown in Fig. 1 (a). It is a binary image when illuminated by a plane wave. In the simulation, we first calculate the field distribution in crystal and then solve the third equation of Eq. (3). The distributions of the flat-top beam and the beam carrying image information are calculated by Collins integral formula [17], where the flat-top field distribution is generated in the focal plane of a lens with the input Airy-disk intensity distribution [14]. And the Gaussian field distribution is directly generated as its analytical form. We set the beam sizes at full width at half maximum (FWHM) of the normalized Gaussian pump and flat-top pump to be the same, which are shown in Fig. 1 (b) and (c). The coupled wave equation is solved by the split-step Fourier method [18]. The solutions are the converted visible field distribution. We then calculate its modular square to obtain the intensity distribution that is the image and shown in Fig. 1 (d) and (e). The size of these images is 0.8mm×0.8mm, and the other parameters are the same as those in the following experiment. The inner products of matrix between the images and the object are 0.9434 and 0.9368 for the flat-top and Gaussian beam, which show the fidelity of frequency up-conversion imaging process. The multiplication in Fig. 1 shows their relation and the inner products show that the flat-top pump can improve the fidelity compared with Gaussian pump.

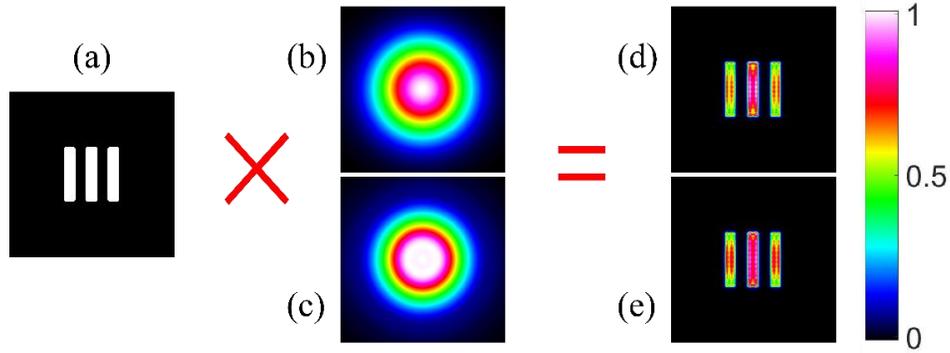

Fig. 1. The simulation of the up-conversion imaging. The multiplication shows the relation between images. (a) The binary object image. (b) Gaussian beam. (c) Flat-top beam. (d) Up-conversion result pumped by Gaussian beam. (e) Up-conversion result pumped by flat-top beam.

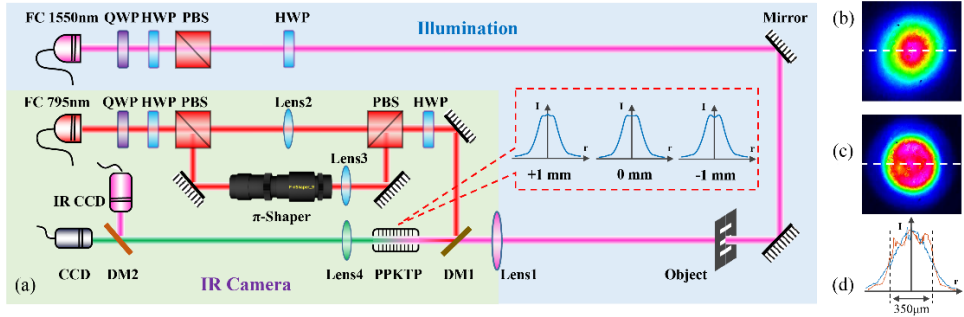

Fig. 2 (a) Experimental set ups for up-conversion imaging system. FC: fiber collimator; QWP: quarter wave plate; HWP: half-wave plate; PBS: polarizing beam splitter; DM: dichroic mirror; The inserts show the three intensity distributions in 2 mm crystal; (b) Gaussian beam intensity distribution in crystal. (c) Flat-top beam intensity distribution in crystal. (d) The one-dimension intensity distribution along the white dotted line in (b) and (c). Blue represents Gaussian beam and orange represents flat-top beam. The color bar is same with that in Fig. 1.

Our experimental setup is displayed in Fig. 2 (a). The 1550 nm illumination beam is provided by a diode laser and is amplified using an erbium-doped fiber amplifier. The pump beam at the wavelength of 795 nm is from a Ti:sapphire laser. Both beams are continuous waves. The half-wave plates (HWPs), the quarter-wave plates (QWPs) and the polarization beam splitters (PBSs) are used in each beam to control their polarization directions for satisfying the phase matching condition. The nonlinear crystal used in our experiment is a type-0 periodically poled potassium titanyl phosphate (PPKTP) crystal with a length of 2 mm and an aperture of 2 mm×1 mm.

In the path through which the infrared beam propagates, a USAF-1951 resolution target is inserted as the object. Lens1 (f=100mm) focuses the infrared light into the crystal, and the center of the crystal overlaps with the image plane. A dichroic mirror (DM1) is used to combine the pump beam and the signal beam containing image information. In the path through which the pump beam propagates, two PBSs are used to separate and combine the paths so that we can easily up-convert images using a Gaussian pump beam or a flat-top pump beam. We rotate the lens on the Pi-shaper to focus the flat-top region at the crystal. Both lens 2 and lens 3 have focal length of 1000 mm. The long focal length allows the pump beam width is larger than the size of the signal image and ensure the flat top beam not diverge in 2 mm distance. And the inserts in fig. 2 (a) show the evolution of flat-top beam in crystal, which is simulated by the focusing Airy-disk intensity distribution [14]. We adjust the position of lens 2 to make the width of the Gaussian beam be the same with the flat-top region at the center of the crystal. Finally, lens 4 (f=100mm) images the up-converted image into the CCD. The distances between the lens 1, the crystal, the lens 4, and the CCD are 750mm, 115mm, 122mm and 550mm, respectively.

We record the pump beam profiles in the crystal imaged by lens 4 and show them in Fig. 2. Fig. 2 (b) and Fig. 2 (c) show the intensity distributions of Gaussian beam and the flat-top beam respectively. Fig. 2 (d) is one dimensional intensity distribution along the white dotted line in Fig. 2 (b) and (c). Blue line represents the Gaussian beam and orange represents the flat-top beam. After we record the beam profile, the positions of CCD and lens 4 are fixed in order to ensure the converted image is pumped by the beam shown in Fig. 2. The IR CCD is helpful to adjust the imaging position of the elements.

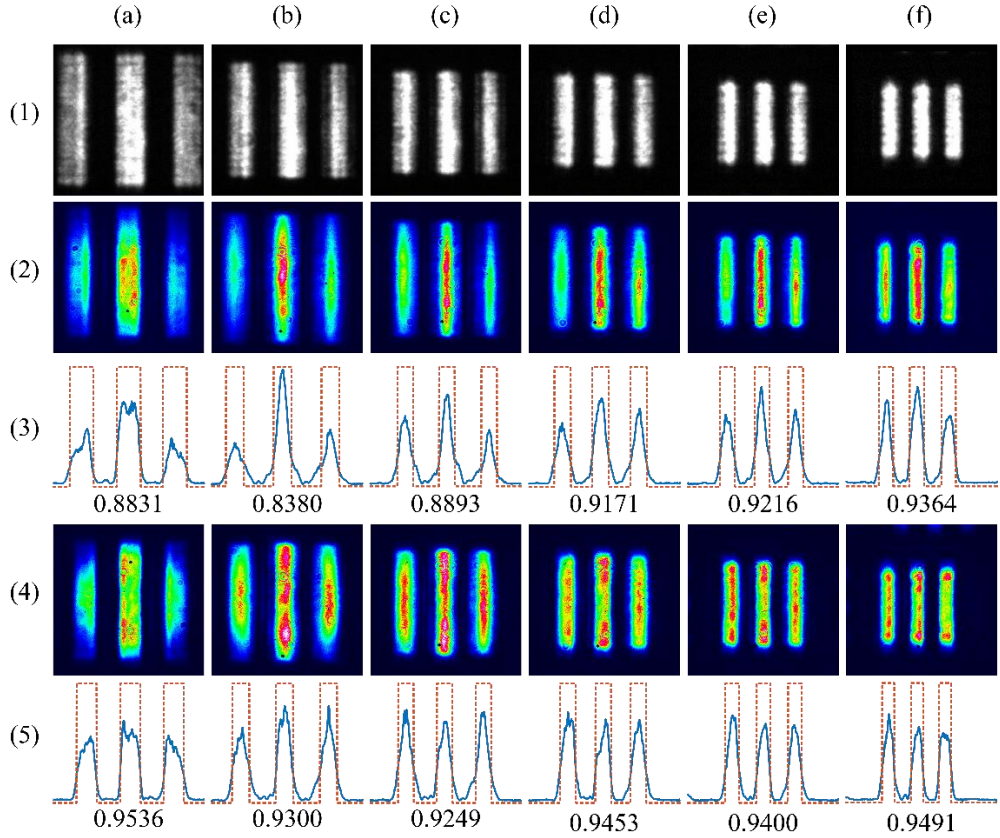

Fig. 3. The experimental result. First row: IR image; The second and the fourth rows: up-converted image pumped by Gaussian and flat-top beam; The third and the fifth rows: the horizontal intensity distribution sampled in the center of the vertical direction and the dotted lines represent theory theoretical distribution; the dotted lines represent the theoretical value; the number is the inner product of experimental and theoretical value. The color bar is same with that in Fig. 1.

We compare the difference between the two pump cases, the results are shown in Fig. 3. The first row shows the IR image; The second and the fourth rows show the up-converted images pumped by Gaussian and flat-top beams respectively; The third and the fifth rows show the horizontal intensity distribution sampled in the center of the vertical direction and the dotted lines represent the ideal intensity distribution. Before sampling the horizontal intensity values from the image data, the image is smoothed by a mean filter to reduce the noise and the fluctuation. The mean filter is a digital image filtering algorithm, which refers to replace every pixel with the average of all the original pixels around it (e.g. the pixels in a 3×3 matrix). The intensities of images are not ideal symmetric because the objects and the beams are not perfectly aligned. The number shown below the one-dimensional intensity distribution is the normalized inner product of experimental results and ideal intensity distribution. We cannot assess the up-conversion imaging quality by comparing with the IR image directly because the IR images and up-converted images are recorded by different CCDs which have different pixel number. The inner product is 1 if the two vectors are the same after normalization. It can be regarded as a criterion to assess the imaging quality in the aspect of the intensity profile. Obviously, the case with a flat-top pump beam is better than that with a Gaussian pump beam. From left to right, when the size of the objects decreases, the two pump schemes become similar. The reason is the center of a Gaussian function is also almost flat. One can also use the Gaussian beam larger than the signal beam to improve the imaging quality. But a larger pump beam leads to low conversion efficiency and wastes the pump power. So in the practical application, one should select the size of a flat-top beam to match the field of view to obtain a better image than a Gaussian pump beam.

We focus on the images (b2) and (b4) in Fig. 3 and further explain the improvement with a flat-top beam pump in the aspect of the application. Usually, image segmentation is an important method in the process of image analysis and is widely used in target recognition. The inhomogeneous intensity distribution makes the segmentation be error-prone. Fig. 4 (a) and (b) show the gray histogram of the images (b2) and (b4) in Fig. 3 respectively. The vertical axis represents number of corresponding pixel in logarithmic coordinate with base-10. We implement the image threshold segmentation by the Matlab function using Otsu's method [19] and show the binary image in Fig. 4 (c) and (d). The value in the histogram of Fig. 4 (a) declines linearly, thus it is hard to select a threshold level to segment the image so that the segmentation image is worse than that pumped by a flat-top beam.

Fig. 4 (e) and (f) show the segmentation images and the raw image is smoothed by a mean filter. We calculate the standard deviation of the segmented region and the results of Fig. 4 (e) and (f) are 0.1552 and 0.1492 respectively. The theoretical standard deviation is zero and the latter is smaller. So the up-converted image pumped by a flat-top beam is more homogeneous and is closer to the image of the object in the aspect of the statistical feature.

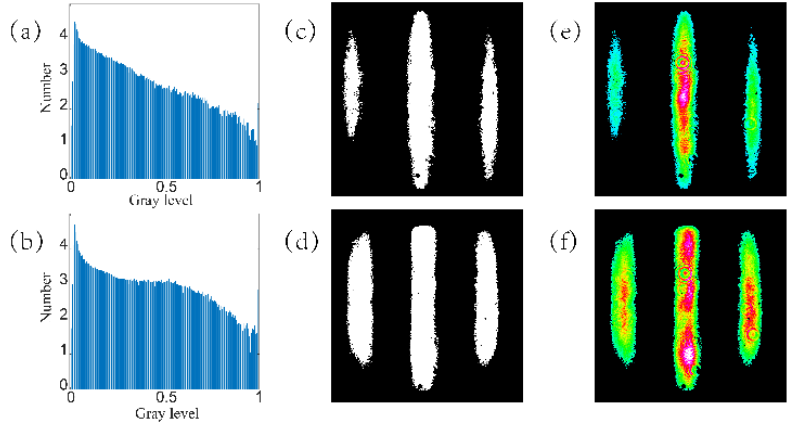

Fig. 4. The processed images of (b2) and (b4) in Fig. 3. (a) and (b): gray histograms of the up-converted images pumped by Gaussian and flat-top beam; The vertical axis represents number of corresponding pixel in logarithmic coordinate with base-10. (c) and (d): the binary image of segmentation region. (e) and (f): the segmentation image. The color bar is the same with that in Fig. 1.

In summary, we study two cases of image up-conversion with the Gaussian and flat-top pump beam. Both the theoretical and experimental results demonstrate the quality of up-converted imaging can be improved with a flat-top beam pump in our image scheme. The current single-pass system works with a low conversion efficiency, one could shape the pulse to generate a flat-top region to improve the imaging conversion efficiency in the future. On the other hand, the beam with a flatter intensity profile has a narrower spatial frequency spectrum, which leads to phase mismatching of the high spatial frequency components in the process of frequency conversion. A pulse pump can not only realize high conversion efficiency but also has broadband which provides high angular acceptance to offset the disadvantage [9]. In addition, the flat-top beam pump method can also be used in the up-conversion of mid-infrared images [10, 11] quantum images [16], and structured beams [20] to improve the operation fidelity.

Another useful method of improving imaging quality is intensity calibration on CCD signals that equalize the intensity distribution of the converted beam. This method is a traditional digital image process, while the flat-top beam pump is a physical process. By comparison, digital processing takes more time, and setting the calibration makes the operation complex and may introduce additional error, while we only need to change the pump beam shape in the physical process. In this present work and our previous work [12], we realize the image up-conversion with intensity equalization via a pump beam with flat intensity and edge enhancement via a pump beam with spiral phase respectively. The common essence is to manipulate the up-converted image by controlling the pump beam properties. We believe the present method is worth being developed for high-quality up-conversion image processing such as digital processing in some situation that requires high-speed processing.

**Funding**